%% file: main.tex
\documentclass[smallextended]{svjour3}

\include{header}

\begin{document}
\title{Optimizing the Decoy-State BB84 QKD Protocol Parameters}

\author{Thomas Attema \and
	Joost W. Bosman \and 
	Niels M. P. Neumann
}

\institute{Thomas Attema
		\at CWI, Cryptology Group, Amsterdam, The Netherlands \and
	      Thomas Attema
		\at Leiden University, Mathematical Institute, Leiden, The Netherlands \and
	      Thomas Attema \and Joost W. Bosman \and Niels M. P. Neumann
		\at TNO, Cyber Security and Robustness, The Hague, The Netherlands \\\email{\{thomas.attema, niels.neumann\}@tno.nl}
}

\date{\today}
\maketitle

\begin{abstract}
	Quantum Key Distribution (QKD) protocols allow for information theoretically secure distribution of (classical) cryptographic key material. 
	However, due to practical limitations the performance of QKD implementations is somewhat restricted. 
	For this reason, it is crucial to find optimal protocol parameters, while guaranteeing information theoretic security. 
	
	The performance of a QKD implementation is determined by the tightness of the underlying security analysis.
	In particular, the security analyses determines the {\em key-rate}, i.e., the amount of cryptographic key material that can be distributed per time unit. 
	Nowadays, the security analyses of various QKD protocols are well understood. 
	It is known that optimal protocol parameters, such as the number of decoy states and their intensities, can be found by solving a nonlinear optimization problem. 
	The complexity of this optimization problem is typically handled by making an number of heuristic assumptions. 
	For instance, the number of decoy states is restricted to only one or two, with one of the decoy intensities set to a fixed value, and vacuum states are ignored as they are assumed to contribute only marginally to the secure key-rate. 
	These assumptions simplify the optimization problem and reduce the size of search space significantly.
	However, they also cause the security analysis to be non-tight, and thereby result in sub-optimal performance. 
	
	In this work, we follow a more rigorous approach using both linear and non-linear programs describing the optimization problem. 
	Our approach, focusing on the Decoy-State BB84 protocol, allows heuristic assumptions to be omitted, and therefore results in a tighter security analysis with better protocol parameters. 
	We show an improved performance for the Decoy-State BB84 QKD protocol, demonstrating that the heuristic assumptions typically made are too restrictive.
	Moreover, our improved optimization frameworks shows that the complexity of the performance optimization problem can also be handled without making heuristic assumptions, even with limited computational resources available. 
		
\end{abstract}

\input{Sections/Introduction.tex}

\input{Sections/BB84.tex}

\input{Sections/Security.tex}
\input{Sections/OptimizingKeyRate.tex}
\input{Sections/Results.tex}
\input{Sections/Conclusion.tex}

\bibliographystyle{ieeetr}
\bibliography{main}

\appendix
\input{Sections/Appendix}

\end{document}

%% file: header.tex
\usepackage[USenglish]{babel}
\usepackage[utf8]{inputenc}
\usepackage{amsmath,amsfonts,amssymb, mathrsfs,mathtools,physics}
\usepackage{graphicx,caption,subcaption,blindtext,float,etoolbox,enumitem,gensymb, fancyhdr,xfrac,makecell,url}
\usepackage[margin=2cm]{caption}
\usepackage{authblk}

\usepackage{todonotes}

\newcommand{\be}{\begin{equation}}
\newcommand{\ee}{\end{equation}}

\newcommand{\Prob}{\text{P}}

\newcommand{\basis}{\mathcal{X}}

%% file: Sections/Introduction.tex
\section{Introduction}

The goal of a key-distribution protocol is for two parties, Alice and Bob, to agree on a key $k\in \{0,1\}^n$ over an insecure communication channel, such that even an adversary Eve with full control over this communication channel can only obtain a negligible amount of information about this key $k$. If Alice and Bob are capable of communicating quantum information, they are able to achieve  information-theoretic or unconditional security, i.e., security against adversaries with unlimited computational power. 

The use of quantum information to securely distribute symmetric cryptographic keys was first proposed in 1984 by Bennett and Brassard~\cite{BB84}, and today their BB84 protocol is still, without doubt, the best-known quantum key distribution (QKD) protocol that exists. Since then, much progress has been made and the first QKD systems are already commercially available. Thereby, QKD has become one of the first applications of quantum mechanics at an individual quanta level~\cite{GRTZ02}. 

The information-theoretic security of the BB84 protocol against the most general attacks allowed by quantum mechanics was proven in 1996 by Mayers~\cite{May96}. In general, however, Mayers' security notion does not imply that the derived key can securely be used in other cryptographic protocols~\cite{BHLMO05} and a stronger notion of security following the universal composability framework is required~\cite{Can01}. Informally, universal security is proven by comparing the output of the protocol to the output of an ideal key-distribution protocol, i.e., a perfect key. If these two outputs are indistinguishable, the protocol is said to be universally secure. Fortunately, QKD protocols that satisfy Mayers' weaker notion of security were shown to be universally secure~\cite{BHLMO05}.

Practical implementations deviate from the theoretical BB84 protocol, which may render them insecure. As any realistic quantum channel introduces noise by imperfections in the source, channel and detector. These benign losses and errors are indistinguishable from the ones introduced by an adversary. For this reason, the conservative assumption is made that all losses and errors are caused by an adversary. Mayers' proof already considered noisy quantum channels, and shows that the BB84 protocol is information-theoretically secure as long as the noise level is below a certain threshold~\cite{May96}. In contrast to the ideal BB84 protocol, practical implementations therefore require an error correction procedure. 
An elaborate review of practical quantum key distribution protocols is given in~\cite{XMZLP20}, including different adversary strategies for practical QKD protocol implementations. 

One of these adversary strategies relates to the photon source. Some protocols, such as the original BB84 protocol, require information to be encoded in single photons. 
However, producing single photon pulses is hard in practice. 
Therefore, typically, the quantum information is encoded in weak laser pulses, where the number of photons in each of these laser pulses follows a Poisson distribution with an attunable mean $\mu$, called the intensity of the source. As a result, laser pulses can contain multiple photons, which can be exploited via the photon-number-splitting (PNS) attack~\cite{BLMS00b,Lut00}. For this reason we must assume that all key material derived from multi-photon pulses is compromised. Privacy amplification is applied to establish a secret key from such a partially compromised key. 

In general, the performance of a QKD implementation can be quantified by the key-rate $R$, indicating the amount of secure key per sent pulse. For the BB84 protocol, this key-rate depends, apart from the noise and losses, also on the laser source intensity $\mu$. By carefully choosing this attunable $\mu$, the key-rate can be maximized. In fact, the protocol does not require the intensity to be fixed throughout the protocol. On the contrary, it has been shown to be beneficial to randomly vary the intensity $\mu$ between pulses, resulting in the so-called decoy-state BB84 protocol with higher achievable key-rates~\cite{Hwa03,LMC05,Wan05}. 

This work focuses on the security analysis of the decoy-state BB84 protocol and the optimization of its protocol parameters. The security analysis is based on the uncertainty relation for smooth entropies~\cite{TR11}. In~\cite{TLGR11}, the uncertainty relation was applied to the analysis of the BB84 protocol implemented with a perfect single photon source. This analysis was extended to the decoy-state BB84 protocol in which weak pulsed laser sources with attunable intensities are applied~\cite{LCWXZ14}. The analysis of~\cite{LCWXZ14} restricts itself to the case where the intensities are randomly varied between three different levels. The intensities and sample distribution are chosen to optimize the key-rate that is achieved in a specific set-up. This approach can be generalized to arbitrary numbers of intensities resulting in a larger parameter search space over which the key-rate is optimized. 

Our main contribution is the formulation of the key-rate optimization problem as linear and non-linear programs.
Analytical lower bounds on the number of multi-photon states can be found~\cite{MQZL05,LCWXZ14}, as the number of photons is assumed to follow a Poisson distribution. Whereas in general these lower bounds are non-tight, the linear programs allow for tight lower bounds, which in turn results in an improved security analysis. Additionally, the linear programs are used to upper bound the number of single-photon errors. Constrained non-linear optimization techniques~\cite{ConnGT00} are then used to optimize the lower bound on the key-rate. 

Due to the larger parameter search space, higher key-rates are found using linear programs, compared to analytical methods. For that reason, linear programs have been used before to optimize the key-rate for the BB84 protocol~\cite{LPD13} and measurement-device independent QKD protocols~\cite{CXC14}. Both works however still make assumptions, for instance by only including the first laser intensity instead of all or by fixing the probabilities for the bases and intensities. 

We formalize the approach and do not make such assumptions. Furthermore, we improve the lower bounds found with the linear programs by including the vacuum and single photon pulses in a single linear program. This opposed to determining lower bounds for the two separately, resulting in conservative and sub-optimal estimates. We furthermore allow for freely varying each of the intensities.  

We show that using this formal approach results in an improved obtainable secure key-rate. We furthermore show the effects of using more decoy states and the effects of increasing the number of sent pulses. First, we explain the BB84 protocol in Section~\ref{sec:BB84} and discuss the security and robustness of the protocol from a mathematical perspective in Section~\ref{sec:CorrectnessRobustness}. The same mathematical perspective is used in Section~\ref{sec:security} to explain how to obtain secure key-material from a BB84 protocol execution. This section also introduces finite key-effects. Afterwards, Section~\ref{sec:model_quantum_channel} explains how to optimize the secure key-rate and how to incorporate the used quantum channel in our model. Results of our model are presented in Section~\ref{sec:Results} and a conclusion is given in Section~\ref{sec:conclusion}.

%% file: Sections/BB84.tex
\section{Decoy-state BB84 protocol}
\label{sec:BB84}
In this section, we recall the Decoy-State BB84 QKD protocol. 
Alice and Bob are assumed to have access to a (noisy) quantum channel and an authenticated classical channel. Both channels are insecure and can be fully controlled by the adversary, however, active attacks on the classical channel are assumed to be immediately detected as this channel is authenticated. 

In the following, all keys are denoted by an uppercase $K$ and the superscripts refer to the party Alice $a$ or Bob $b$ and type of key: raw $r$, sifted $s$ or error-corrected $e$. 
The decoy-state BB84 protocol now goes as follows. 

\textbf{Preparation:} Alice generates a raw key by sampling a uniformly random bit string in $K^{ra}\in \{0,1\}^N$. Moreover, Alice samples a random basis string $\basis^a \in \{X,Z\}^N$, where $P(\basis^a_i=X)=p_X$ for all $1\leq i \leq N$. For all $i$ she encodes bit $K^{ra}_i$ in basis $\basis^a_i$ resulting in a sequence of qubits in $\left\{\ket{0},\ket{1},\ket{+},\ket{-} \right\}^N$.

\textbf{Communication:} For all $1 \leq i \leq N$, Alice samples an intensity $U_i$ from a probability distribution $P_{U_i|\basis^a_i}$ conditioned on the chosen basis $\basis_i^a \in \{X,Z\}$. The distribution $P_{U_i|\basis^a_i}$ is independent of the index $i$ and for both bases its support lies in a finite set of intensities $\{\mu_0,\dots,\mu_m \}$. Alice encodes the associated qubit in a laser pulse with intensity $U_i$ and sends this pulse to Bob over the quantum channel. 

\textbf{Measurement:} Bob samples a random basis string $\basis^b \in \{X,Z\}^N$, where $P(\basis^b_i=X)=p_X$ for all $1\leq i \leq N$ and measures pulse $i$ in basis $\basis^b_i$. If both of Bob's detectors register an event, for example in the case of a multi-photon pulse and a measurement incompatible with Alice's preparation, Bob randomly selects a measurement outcome $K^{rb}_i \in \{0,1 \}$. It can also occur that no detector registers an event, in this case Bob defines the outcome to be $\emptyset$. As a result Bob obtains his raw key $K^{rb} \in \{0,1,\emptyset\}^N$. Note that Bob's sifting probability $p_X^b$ is equal to that of Alice $p_X^a$, i.e., $p_X^a=p_X^b=p_X$. This is not required, but it can be shown that for all protocol instantiations with $p_X^a \neq p_X^b$, there exists a protocol instance with $p_X^a=p_X^b$ with at least the same secure key-rate.

\textbf{Sifting:} Alice and Bob announce their basis choices and Bob announces the pulses for which no detection event took place. The pulses that were prepared and measured in the same basis and for which a detection event occurred, are sifted from the raw keys $K^{ra}$ and $K^{rb}$  and Alice and Bob obtain sifted keys $K^{sa}\in \{0,1\}^{n_s}$ and $K^{sb} \in \{0,1\}^{n_s}$ respectively. In addition, we let $K^{sa}_{\basis},K^{sb}_{\basis}\in \{0,1\}^{n_{\basis}}$ be the strings containing the bits of $K^{sa}$ and $K^{sb}$ obtained by preparing and measuring in the $\basis$-basis for $\basis\in\{X,Z\}$.

\textbf{Parameter estimation:} Alice announces the chosen intensities $U$, which allows Alice and Bob to compute the amount of detection events 
\begin{equation}
\label{eq:nr_events}
\begin{split}
n_{\mu_j,\basis} = \left\lvert \left\{i: \basis^a_i = \basis^b_i=\basis \land U_i=\mu_j  \land K^{rb}_i \neq \emptyset \right\} \right\rvert,
\end{split}
\end{equation}
for all intensities $\mu_j$ and for both bases $\basis \in \{X,Z\}$. In addition, Alice and Bob reveal the parts of the sifted keys $K^{sa}_Z,K^{sb}_Z\in \{0,1\}^{n_Z}$. Using this information they can compute the amount of errors in the $Z$-basis
\begin{equation}
\label{eq:nr_errors}
\begin{split}
E_{\mu_j,Z} = \left\lvert \left\{i: \basis^a_i = \basis^b_i=Z \land U_i=\mu_j \land (K^{sa}_Z)_i \neq (K^{sb}_Z)_i \right\}  \right\rvert,
\end{split}
\end{equation}
for all intensities $\mu_j$. Given these values Alice and Bob determine upper-bounds on the number of bits in $K^{sa}_X$ that are associated to multi-photon events and the error-rate $e_{1,Z}$ for single photon pulses in the $Z$-basis. From these bounds Alice and Bob determine the length $\ell$ of the secret key that can be extracted after the error reconciliation phase. If $\ell\leq 0$ the protocol aborts. Note that only the $Z$-events are used to determine $\ell$. The $X$-events will be used in the error reconciliation and privacy amplification phase to construct the final key. 

\textbf{Error reconciliation:} Errors in the quantum channel can cause the strings $K^{sa}_X$ and $K^{sb}_X$ to be distinct. For this reason Alice and Bob perform an information reconciliation protocol by which they obtain error-corrected keys $K^{ea},K^{eb}\in \{0,1\}^{n_X}$ respectively. 

\textbf{Verification:} Alice samples a uniformly random hash function $h$ from a two-universal family of hash functions $\mathcal{F}_e:\{0,1\}^{n_{X}} \rightarrow \{0,1\}^{\left\lceil -\log_2(\epsilon_{cor}) \right\rceil}$~\cite{CW79}. Here $0\leq 1-\epsilon_{cor} \leq 1$ is a lower bound on the probability that the protocol is correct, i.e., that Alice and Bob will obtain identical keys. Alice applies this hash function to her error-corrected key $K^{ea}$. She sends the hash-function $h$ and hash-value $h(K^{ea})$ to Bob, who then computes $h(K^{eb})$. If $h(K^{ea}) \neq h(K^{eb})$, the protocol aborts. 

\textbf{Privacy Amplification:} Alice samples a uniformly random hash function $h$ from a two-universal family $\mathcal{F}_p$ mapping  $\{0,1\}^{n_X}$ to $\{0,1\}^{\ell}$ and announces $h$ to Bob, where $\ell$ has been determined in the parameter estimation phase. Both Alice and Bob compute the secret keys $K^a = h(K^{ea})$ and $K^b=h(K^{eb})$ respectively.

%% file: Sections/Security.tex
\section{Correctness and Robustness}
\label{sec:CorrectnessRobustness}

In this section, we recall two important (security) properties of QKD protocols. 
A QKD protocol should be correct and secure against any attack allowed by quantum mechanics. Moreover, the protocol should only abort with a small probability, i.e., it should be robust. In this section we formalize the correctness and robustness properties, following the approach of~\cite{Renner05}, and show why the decoy-state BB84 protocol admits these properties. The security of the protocol will be analyzed in Section~\ref{sec:security}. 

A QKD protocol is $\epsilon_{\text{cor}}$-correct if 
\begin{equation}
P(K^a \neq K^b) \leq \epsilon_{\text{cor}},
\end{equation}
for all possible strategies of an adversary. This property is easily seen to be satisfied if in the verification phase $\mathcal{F}_e$ is indeed a family of two-universal hash functions mapping $\{0,1\}^{n_X}$ to $\{0,1\}^{\left\lceil - \log_2 \left(\epsilon_{\text{cor}} \right) \right\rceil}$. The hash value reveals 
\begin{equation}
\label{eq:log_epsCorr}
\left\lceil\log_2\left(\frac{1}{\epsilon_{\text{cor}}}\right)\right\rceil \leq \log_2\left(\frac{2}{\epsilon_{\text{cor}}}\right)
\end{equation} 
bits of information to the adversary. 

Note that the protocol is allowed to abort, in which case $K^a = K^b=\bot$. The probability $p_{\text{abort}}$ that the protocol aborts in the absence of an adversary, depends on the error reconciliation protocol that is applied. For any $\delta_{\text{ec}}> 0$, there exist error reconciliation protocols that leak at most 
\begin{equation}
\label{eq:numEvents_X}
n_X \left( h(e_X) + \delta_{\text{ec}} \right) 
\end{equation}
bits of information~\cite{Renner05}, where $h$ is the binary entropy function and $e_X$ is the quantum bit error rate (QBER) of the sifted keys in the $X$-basis. Moreover, 
\begin{equation}
\begin{split}
p_{\text{abort}} & \leq \Prob \left( K^{ea} \neq K^{eb} \right)  \leq 2 \exp\left(\frac{-n_X\delta_{\text{ec}}^2}{3\log_2(5)^2}\right), 
\end{split}
\end{equation}
where the last inequality follows from Corollary~$6.3.5$ of~\cite{Renner05}. To achieve an abort probability of at most $p_{\text{abort}}$ we therefore take 
\begin{equation}
\delta_{\text{ec}}(p_{\text{abort}},n_X) = \sqrt{\ln\left(\frac{2}{p_{\text{abort}}}\right)\frac{3\log_2(5)^2}{n_X}}.
\end{equation} 

\section{Security of the Decoy-State BB84 Protocol }
\label{sec:security}

In this section, we recall the standard (composable) security definitions for QKD protocols, specifically focusing on the Decoy-State BB84 protocol. 
In particular, we derive an expression for the length of a key that can securely be generated by running this QKD protocol (Section~\ref{sec:sec_key_length}). 
This expression contains a number of variables that are unknown to Alice and Bob. In Section~\ref{sec:unknown}, we show how these unknown protocol variables can be bounded by solving certain linear programs. 

\subsection{Secure Key Length}
\label{sec:sec_key_length}

To evaluate the security of the protocol let us consider the joint state of the classical random variable $K:=K^a$ with support $\mathcal{K}$ and the adversary's quantum system 
\begin{equation}
\rho_{KE} = \sum_{x\in \mathcal{K}} P(K=x) \ket{x}\bra{x} \otimes \rho_E^x,
\end{equation} 
where $\rho_E^x$ is the state of the adversary's system given that $K=x$. Ideally, the classical probability distribution $P(K=x)$ is uniform and the adversary's state is independent of $K$. Hence, the joint state of a perfect key and the adversary's system is given by 
\begin{equation}
\frac{1}{\left\lvert \mathcal{K} \right\rvert}\sum_{x\in\mathcal{K}} \ket{x}\bra{x}  \otimes \rho_E.
\end{equation} 
A QKD-protocol is now said to be $\epsilon_{\text{sec}}$-secure if 
\begin{equation}
\frac{1}{2} \left\lVert \rho_{KE} - \frac{1}{\left\lvert \mathcal{K} \right\rvert}\sum_{x\in\mathcal{K}} \ket{x}\bra{x}  \otimes \rho_E \right\rVert_1 \leq \epsilon_{\text{sec}},
\end{equation}
where $\lVert A \rVert_1 = \tr \left( \sqrt{A^* A} \right)$ is the trace norm of the complex matrix $A$. 

If a QKD-protocol is $\epsilon_{\text{sec}}$-secure the output cannot be distinguished from that of a perfect protocol with probability more than $\epsilon_{\text{sec}}$~\cite{Renner05}. Moreover this security definition ensures universal composability, i.e., the key $K$ can safely be used in other cryptographic protocols. 

In general, Alice's bit-string $K^{ea}\in \{0,1\}^{n_X}$, obtained after the error correction and verification phase, does not satisfy the above security definition. For this reason privacy amplification is applied. From the leftover hash lemma~\cite{TSSR11} it follows that the QKD protocol is $\epsilon_{\text{sec}}$-secure if there exists an $\epsilon\geq 0$ such that
\begin{equation}
\label{eq:rate}
\ell = \left\lfloor  H_{\min}^{\epsilon}(K^{ea}|E) - 2 \log_2 \left(\frac{1}{2(\epsilon_{\text{sec}}-\epsilon)} \right) \right\rfloor,
\end{equation}
where $H_{\min}^{\epsilon}(K^{ea}|E)$ is the smooth min-entropy of the random variable $K^{ea}$ conditioned on the adversary's (quantum) information $E$. The leftover hash lemma thus gives an expression of the bit-length $\ell$ of the secure key $K$ in terms of this conditional smooth entropy. 

The error corrected key is obtained from the sifted key after performing the error correction and verification phase, which both leak some information to the adversary. If we let $E^s$ be the adversary's (quantum) information after the sifting phase, it follows from Equations~\eqref{eq:log_epsCorr} and~\eqref{eq:numEvents_X} that
\begin{equation}
\label{eq:error}
\begin{split}
H_{\min}^{\epsilon}(K^{ea}|E) \geq & H_{\min}^{\epsilon}(K^{sa}_X|E^s) - n_X \left( h(E_X)+\delta_{\text{ec}}(p_{\text{abort}},n_X)\right) - \log_2\left(\frac{2}{\epsilon_{\text{cor}}}\right) . 
\end{split}
\end{equation}

Now observe that the bits of $K^{sa}_X$ are all derived from vacuum, single-photon or multi-photon pulses. Hence, we can write
\begin{equation}
K^{sa}_X=K^{sa}_{0,X} \otimes K^{sa}_{1,X} \otimes K^{sa}_{m,X},
\end{equation}
where $K^{sa}_{0,X}$, $K^{sa}_{1,X}$ and $K^{sa}_{m,X}$ contain the bits associated to the vacuum, single-photon and multi-photon pulses, respectively. 

It is impossible for an adversary to obtain information about the bits associated to vacuum pulses, hence for all $\epsilon\geq 0$ 
\begin{equation}
\label{eq:vacuum}
H^{\epsilon}_{\min}(K^{sa}_{0,X}|E^s) \geq H_{\min}(K^{sa}_{0,X}|E^s) = n_{0,X},
\end{equation}
where $n_{0,X}$ is the number of vacuum pulses that were sent. In contrast, the PNS attack allows the adversary to obtain all information about the bits associated to multi-photon pulses. Hence, for all $\epsilon\geq 0$ we must lower bound the associated min-entropy as follows
\begin{equation}
\label{eq:multi}
H^{\epsilon}_{\min}(K^{sa}_{m,X}|K^{sa}_{0,X}E^s) \geq 0. 
\end{equation}

Applying the chain rule for smooth min-entropies~\cite{VDTR13} twice and plugging in Equations~\eqref{eq:vacuum} and~\eqref{eq:multi}, we find that for all $\epsilon,\epsilon_1,\epsilon_4,\epsilon_5\geq 0$ and $\epsilon_2, \epsilon_3>0$ such that $2\epsilon_1 +  \epsilon_2 + \epsilon_3 + 2\epsilon_4 + \epsilon_5 = \epsilon \leq 1$,
\begin{equation}
\label{eq:chain}
\begin{split}
H_{\min}^{\epsilon}(K^{sa}_X|E^s) & \geq  H^{\epsilon_5}_{\min}(K^{sa}_{0,X}|E^s) + H_{\min}^{\epsilon_1}(K^{sa}_{1,X}|\tilde{E}) + H^{\epsilon_4}_{\min}(K^{sa}_{m,X}|K^{sa}_{0,X}E^s)  \\&\quad  - \log_2\left(\frac{1}{1-\sqrt{1-\epsilon_2^2}}\right) - \log_2\left(\frac{1}{1-\sqrt{1-\epsilon_3^2}}\right) \\
&\geq n_{0,X} + H_{\min}^{\epsilon_1}(K^{sa}_{1,X}|\tilde{E}) \\& \quad - \log_2\left(\frac{1}{1-\sqrt{1-\epsilon_2^2}}\right) - \log_2\left(\frac{1}{1-\sqrt{1-\epsilon_3^2}}\right) \\
&\geq n_{0,X} + H_{\min}^{\epsilon_1}(K^{sa}_{1,X}|\tilde{E})  - 2\log_2\left(\frac{2}{\epsilon_2\epsilon_3}\right),
\end{split}
\end{equation}
where $\tilde{E} = K^{sa}_{0,X} \otimes K^{sa}_{m,X}\otimes E^s$ and for the third inequality we use that for all $x\leq 1$ it holds that 
\[
1-\sqrt{1-x} \geq \frac{x}{2}.
\]
See also~\cite{LCWXZ14} in which the same lower bound is derived. 

Hence, combining Equations~\eqref{eq:rate},~\eqref{eq:error} and~\eqref{eq:chain} gives an achievable secure key-rate in terms of the smooth min-entropy of the $n_{1,X}$ single-photon pulses in the $X$-basis conditioned on the quantum system $\tilde{E}$. For these pulses we have the following uncertainty relation~\cite{TR11}, 
\begin{equation}
\begin{split}
H_{\min}^{\epsilon_1}(K^{sa}_{1,X}|\tilde{E})  \geq n_{1,X} - H_{\max}^{\epsilon_1}(L^{sa}_{1,Z}|L^{sb}_{1,Z}),
\end{split}
\end{equation}
where $L^{sa}_{1,Z}$ and $L^{sb}_{1,Z}$ are the hypothetical sifted keys that would have been obtained if Alice and Bob would have prepared and measured the $K^{sa}_{1,X}$-pulses in the $Z$-basis. Informally, this uncertainty relation states that either Eve is uncertain about Alice's key in the $X$-basis or Alice and Bob observe a high amount of errors in their $Z$-events. 

Let $e_{1,Z}^L$ be the fraction of errors between $L^{sa}_{1,Z}$ and $L^{sb}_{1,Z}$ and let $e_{1,Z}$ be the fraction of errors between $K^{sa}_{1,Z}$ and $K^{sb}_{1,Z}$. The total fraction of single-photon errors that would have been obtained if Alice and Bob had prepared and measured all these pulses in the $Z$-basis then equals
\begin{equation}
e_{1,Z}^{\text{tot}} = \frac{n_X e_{1,Z}^L + n_Z e_{1,Z}}{n_X+n_Z}. 
\end{equation}
The amount of errors $n_Z e_{1,Z}$ can now be seen to be equal to the number of errors in a subset of size $n_Z$ randomly sampled from a set of size $n_X+n_Z$ containing $(n_X+n_Z)e_{1,Z}^{\text{tot}}$ errors. Hence, $n_Z e_{1,Z}$ follows a hypergeometric distribution and 
\begin{equation}
\label{eq:hyper_tail}
\begin{split}
\Prob\left(e_{1,Z}^L \geq e_{1,Z} + \delta\right) & = \Prob\left(n_Ze_{1,Z} \leq n_Z e_{1,Z}^{\text{tot}} - \frac{n_Xn_Z\delta}{n_X+n_Z} \right), \\
& \leq \exp\left(\frac{-2n_X^2n_Z\delta^2}{\left(n_X+n_Z\right)\left(n_X+1\right)}\right), \\
\end{split}
\end{equation}
where the inequality follows by applying Serfling's tail bound of the hypergeometric distribution~\cite{Serfling1974}. This upper bound is slightly different from that of~\cite{TLGR11}. If $n_X \geq n_Z$, then Equation~\eqref{eq:hyper_tail} gives a tighter upper bound than~\cite{TLGR11}, but the difference between the two bounds is negligible. By Equation~\eqref{eq:hyper_tail} the event that an adversary correctly guesses the basis choices is taken into account for example. 

If we now take 
\begin{equation}
\begin{split}
\delta = \delta(n_X,n_Z,\epsilon_1) = \sqrt{\frac{(n_X+n_Z)(n_X+1)}{2n_X^2n_Z} \ln\left(\frac{1}{\epsilon_1}\right)},
\end{split}
\end{equation}
we find that $\Prob\left(e_{1,Z}^L \geq e_{1,Z} + \delta\right) \leq \epsilon_1$. It now follows that 
\begin{equation}
\label{eq:hypothetical_key_entropy}
\begin{split}
H_{\max}^{\epsilon_1}(L^{sa}_{1,Z}|L^{sb}_{1,Z}) \leq n_{1,X} \bar{h}\left( e_{1,Z}+\delta(n_X,n_Z,\epsilon_1)\right),
\end{split}
\end{equation}
were $\bar{h}(p)=h\left(\min\left(p,1/2\right)\right)$ for the binary entropy function $h$~(Lemma~3 of~\cite{TLGR11}). Note that in the asymptotic limit, i.e., $n_X,n_Z\to \infty$, the term $\delta$ vanishes. 

Altogether, we thus find that the QKD protocol is $\epsilon_{\text{sec}}$-secure if 
\begin{equation}
\label{eq:key-length}
\begin{split}
\ell  = & \biggl\lfloor n_{0,X} + n_{1,X} - n_{1,X}\bar{h}\left( e_{1,Z}+\delta(n_X,n_Z,\epsilon_1)\right) \\ & -  n_X\left(h(e_X)+\delta_{\text{ec}}\left(p_{\text{abort}},n_X\right)\right) -   \log_2\left(\frac{2}{\epsilon_{\text{cor}}(\epsilon_2\epsilon_3(\epsilon_{\text{sec}}-\epsilon))^2}\right)\biggr\rfloor, 
\end{split}
\end{equation}
for some $\epsilon_1,\epsilon_2,\epsilon_3>0$ and $\epsilon = 2\epsilon_1+\epsilon_2+\epsilon_3$. The values $\epsilon,\epsilon_2,\epsilon_3$ can be chosen to maximize the length $\ell$ of the secure key. Note that this key-length is conditioned on the fact that the protocol does not abort. The expected key rate of the protocol is therefore given by,
\begin{equation}
\label{eq:key-rate}
\begin{split}
R  = & \frac{1-p_{abort}}{N}\biggl\lfloor n_{0,X} + n_{1,X} - n_{1,X}\bar{h}\left( e_{1,Z}+\delta(n_X,n_Z,\epsilon_1)\right) \\ & -  n_X\left(h(e_X)+\delta_{\text{ec}}\left(p_{\text{abort}},n_X\right)\right) -   \log_2\left(\frac{2}{\epsilon_{\text{cor}}(\epsilon_2\epsilon_3(\epsilon_{\text{sec}}-\epsilon))^2}\right)\biggr\rfloor, 
\end{split}
\end{equation}

\subsection{Linear Programs to Bound the Unknown Parameters}
\label{sec:unknown}

Some of the parameters, such as the amount of successful detection events $n_X$ in the $X$-basis, in the key rate expression of Equation~\ref{eq:key-rate} can be observed by Alice and Bob during the execution of the QKD protocol. However, Alice and Bob remain oblivious to other parameter values in this expression. For instance, we assume that Alice and Bob can not distinguish between single and multi-photon events. For this reason, they can not determine the number of single photon events $n_{X,1}$ in the $X$-basis.  
To this end, Alice and Bob resort to upper and lower bounds for these unknown parameter values.
In this section we describe the linear programs that are used to find these bounds. 

Let us first consider the different parameters of Equation~\ref{eq:key-rate}.
The amount of successful detections $n_X$ in the $X$-basis and the QBER $e_X$ can be observed by Alice and Bob. The QBER $e_X$ can be estimated before running the QKD protocol, hence no key material has to be sacrificed to estimate this value. A different QBER during the QKD protocol, possibly due to adversarial behavior, will not compromise the security, but merely influence the abort probability of the protocol. Note that the adversary, controlling the quantum channel, is always capable of aborting the protocol, i.e., performing a denial-of-service attack. Since Alice and Bob are unable to determine the amount of photons per pulse, the variables $n_{0,X}$, $n_{1,X}$ and $e_{1,X}$ remain unknown. For this reason the observable quantities $n_{\mu_j,\basis}$ (Equation~\eqref{eq:nr_events}) and $E_{\mu_j,Z}$ (Equation~\eqref{eq:nr_errors}) for all intensities $\mu_j$ and both bases $\basis$ are used to upper bound the error rate $e_{1,Z}$ and to lower bound the expression $n_{0,X} + n_{1,X} - n_{1,X}\bar{h}\left( e_{1,Z}+\delta(n_X,n_Z,\epsilon_1)\right)$. These bounds result in lower bounds on the secure key length $\ell$. 

Let $n_{l,\basis}$ be the amount of $l$-photon pulses detected by Bob and prepared and measured in the $\basis$-basis. Then, for $0 \leq j \leq m$, it holds that the expected number of $\basis$-pulses sent with intensity $\mu_j$ equals
\begin{equation}
\mathbb{E}\left[n_{\mu_j,\basis}\right] = \sum_{l=0}^{\infty} p_{\mu_j|l,\basis}n_{l,\basis},
\end{equation}
where $p_{\mu_j|l,\basis}$ is the probability that an $l$-photon pulse is sent with intensity $\mu_j$ given that Alice and Bob chose basis $\basis$. Since the amount of photons in a weak laser pulse follows a Poisson distribution, we find by Bayes' rule that,
\begin{equation}
\label{eq:constraint1}
p_{\mu_j|l,\basis}=  \frac{e^{-\mu_j}\mu_j^{l}}{l!} \frac{p_{\mu_j|\basis}}{p_{l|\basis}},
\end{equation}
where $p_{\mu_j|\basis}$ is the probability that an $\basis$-pulse is sent with intensity $\mu_j$ and 
\begin{equation}
p_{l|\basis} =  \sum_{j=0}^m \frac{p_{\mu_j|\basis}e^{-\mu_j} \mu_j^l}{l!},
\end{equation}
is the probability that an $\basis$-pulse consists of $l$ photons~\cite{MQZL05,LCWXZ14}. 

In addition, the number of $l$-photon pulses in the $\basis$ basis $n_{l,\basis}$ that result in a detection event is upper bounded by the number of $l$-photon pulses $N_{l,\basis}$ sent by Alice and measured by Bob in the $\basis$-basis. Note that we use an uppercase $N$ to denote the amount of pulses sent by Alice and a lowercase $n$ to denote the number of events detected by Bob. Hence, 
\begin{equation}
\label{eq:constraint2}
\begin{split}
n_{l,\basis} \leq N_{l,\basis} & \quad \text{and} \quad \mathbb{E}\left[N_{l,\basis}\right] = p_{l|\basis} N_{\basis},
\end{split}
\end{equation}
for all $l\geq0$ and for all $\basis \in \{X,Z\}$. 

Alice and Bob cannot exactly determine the values $n_{l,\basis}$, but Equations~\eqref{eq:constraint1} and~\eqref{eq:constraint2} do supply them with constraints on these values. The variables $n_{\mu_j,\basis}$ are measured in the parameter estimation phase. In the asymptotic limit these estimations are equal to their expected values. Hence, neglecting finite key effects, we can find a lower bound $n_{1,Z}^*$ for $n_{1,Z}$ by solving the following linear program over the unknown variables $n_{l,Z}$ for $l\geq 0$. 

\begin{equation}
\boxed{
\begin{aligned}
n_{1,Z}^*  = &&& \min & & n_{1,Z}, & \\
&&& \text{s.t.} & &  \forall \, 0 \leq j \leq m, \forall \, l \geq 0 \\
&&&&& n_{\mu_j,Z} = e^{-\mu_j}p_{\mu_j|Z} \sum_{l=0}^{\infty} \frac{\mu_j^{l}}{l!} \frac{n_{l,Z}}{p_{l|Z}}, \\
&&&&& 0 \leq n_{l,Z} \leq p_{l|Z}N_Z.
\end{aligned}
}
\end{equation}

However, in all practical situations the amount of pulses is finite and the finite key effects cannot be neglected. By Hoeffding's bounds~\cite{Hoeffding63} we find that 
\begin{equation}
\Prob\left(\left\lvert n_{\mu_j,\basis}- \mathbb{E}\left[n_{\mu_j,\basis}\right] \right\rvert \geq \sqrt{-\ln(\epsilon^H_{\mu_j}/2)n_{\basis}/2} \right) \leq \epsilon^H_{\mu_j,\basis}, 
\end{equation}
for all $0\leq j \leq m$, $\basis \in \{X,Z\}$ and $\epsilon^H_{\mu_j}>0$. Since the variables $N_{l,\basis}$ are sums of Bernoulli random variables we can apply Chernoff's bounds~\cite{Chernoff1952}. In this effort, let us define the following function,
\begin{equation}
f: \mathbb{Z}_{\geq 0} \times [0,1] \times (0,1] \to \mathbb{R}_{\geq 0} \to , \quad (N,p,\epsilon) \mapsto -\ln(\epsilon)\left(1+\sqrt{1-\frac{2pN}{\ln(\epsilon)}}\right).
\end{equation}
Then by Chernoff's bound,  
\begin{equation}
\Prob\left(N_{l,\basis} \geq \mathbb{E}\left[N_{l,\basis}\right] + f(N_{\basis};p_{l|\basis};\epsilon^C_{l,\basis})\right) \leq \epsilon^C_{l,\basis}, 
\end{equation}
for all $l\geq0$, $\basis \in \{X,Z\}$ and $\epsilon^C_{l,Z}>0$. 

Hence, a lower bound $n_{1,Z}^*$ of $n_{1,Z}$, that holds except with probability at most $\sum_{l = 0}^{\infty} \epsilon^C_{l,Z} + \sum_{j=0}^m\epsilon^H_{\mu_j,Z}$, is given by the linear program of Equation~\eqref{eq:LP_Z}.

\begin{equation}
\label{eq:LP_Z}
\boxed{
\begin{aligned}
n_{1,Z}^*  = &&& \min & & n_{1,Z}, & \\
&&& \text{s.t.} & & \forall \, 0 \leq j \leq m, \forall \, l \geq 0 \\
&&&&& n_{l,z}, \delta_{\mu_j} \in \mathbb{R}, \\
&&&&& n_{\mu_j,Z} + \delta_{\mu_j} = e^{-\mu_j}p_{\mu_j|Z} \sum_{l=0}^{\infty} \frac{\mu_j^{l}}{l!}\frac{n_{l,Z}}{p_{l|Z}}, \\
&&&&&  \sum_{j=0}^m \delta_{\mu_j} =0, \\
&&&&& 0 \leq n_{l,Z} \leq \min\left(p_{l|Z}N_Z+f(N_Z;p_{l|Z};\epsilon^C_{l,Z}), n_Z \right), \\ 
&&&&&  \left\lvert \delta_{\mu_j} \right\rvert \leq \sqrt{-\ln(\epsilon^H_{\mu_j,Z}/2)n_Z/2}.
\end{aligned}
}
\end{equation}

Similarly, an upper bound $E_{1,Z}^*$, that holds except with probability at most $\sum_{l = 0}^{\infty} \epsilon^C_{l,Z} + \sum_{j=0}^m\epsilon^H_{\mu_j,E}$, of the number of errors in single-photon $Z$-pulses $E_{1,Z}$ is found by solving the linear program of Equation~\eqref{eq:LP_E}.

\begin{equation}
\label{eq:LP_E}
\boxed{
\begin{aligned}
E_{1,Z}^* = &&& \max & & E_{1,Z} & \\
&&& \text{s.t.} & & \forall \, 0 \leq j \leq m, \forall \, l \geq 0  \\
&&&&& E_{l,z}, \delta_{\mu_j} \in \mathbb{R}, \\
&&&&& E_{\mu_j,Z} + \delta_{\mu_j} = e^{-\mu_j}p_{\mu_j|Z} \sum_{l=0}^{\infty} \frac{\mu_j^{l}}{l!} \frac{E_{l,Z}}{p_{l|Z}}, \\
&&&&&  \sum_{j=0}^m \delta_{\mu_j} =0, \\
&&&&& 0 \leq E_{l,Z} \leq \min\left(p_{l|Z}N_{Z} + f(N_Z;p_{l|Z};\epsilon^C_{l,Z}), E_Z \right),  \\
&&&&&  \left\lvert \delta_{\mu_j} \right\rvert \leq \sqrt{-\ln(\epsilon^H_{\mu_j,E}/2)E_Z/2}.
\end{aligned}
}
\end{equation}

It follows that 
\begin{equation}
e_{1,Z} \leq e_{1,Z}^* = \min \left( \frac{E_{1,Z}^*}{n_{1,Z}^*}, \frac{1}{2}\right)
\end{equation}
except with probability at most
\begin{equation}
\epsilon_e = \sum_{l = 0}^{\infty} \epsilon^C_{l,Z} + \sum_{j=0}^m \left( \epsilon^H_{\mu_j,Z} + \epsilon^H_{\mu_j,E}\right).
\end{equation}

Finally, a lower bound $n_{0,1,X}^*$ of the expression 
\begin{equation} 
n_{0,X} + n_{1,X} - n_{1,X}\bar{h}\left( e_{1,Z}^*+\delta(n_X,n_Z,\epsilon_1)\right),
\end{equation} 
that holds except with probability at most $\epsilon_X = \sum_{l = 0}^{\infty} \epsilon^C_{l,X} + \sum_{j=0}^m\epsilon^H_{\mu_j,X}$, is found by solving the linear program of Equation~\eqref{eq:LP_X}. 
The contribution of the vacuum states, $n_{0,X}$, to the secure key-rate can be argued to be marginal. 
For this reason, the optimization problem of Equation~\eqref{eq:LP_X} is often simplified by ignoring the $n_{0,X}$ component in the objective function. 

\begin{equation}
\label{eq:LP_X}
\boxed{
\begin{aligned}
n_{0,1,X}^* = &&& \min & & n_{0,X}+n_{1,X}-n_{1,X}h(e_{1,Z}^*+\delta(n_X,n_Z,\epsilon_1)), & \\
&&& \text{s.t.} & & \forall \,0 \leq j \leq m, \forall \,l \geq 0 \\
&&&&& n_{l,X}, \delta_{\mu_j} \in \mathbb{R}, \\
&&&&& n_{\mu_j,X} + \delta_{\mu_j} = e^{-\mu_j}p_{\mu_j|X} \sum_{l=0}^{\infty} \frac{\mu_j^{l}}{l!} \frac{n_{l,X}}{p_{l|X}},  \\
&&&&&  \sum_{j=0}^m \delta_{\mu_j} =0, \\
&&&&& 0 \leq n_{l,X} \leq  \min\left(p_{l|X}N_X+f(N_X;p_{l|X};\epsilon^C_{l,X}), n_X \right),  \\
&&&&&  \left\lvert \delta_{\mu_j} \right\rvert \leq \sqrt{-\ln(\epsilon^H_{\mu_j,X}/2)n_X/2}.
\end{aligned}
}
\end{equation}

A detail that has been omitted so far is the fact that these linear programs contain an infinite number of variables, which can be dealt with by truncating the infinite sums at $M$. For the resulting linear programs, with a finite number of variables, we refer to Appendix~\ref{ap:finite_vars}. The same truncation applies to the error terms $\epsilon_X$ and $\epsilon_e$, i.e., 
\begin{equation}
\begin{split}
\epsilon_X & = \sum_{l = 0}^{M} \epsilon^C_{l,X} + \sum_{j=0}^m\epsilon^H_{\mu_j,X}, \\
\epsilon_e & = \sum_{l = 0}^{M} \epsilon^C_{l,Z} + \sum_{j=0}^m \left( \epsilon^H_{\mu_j,Z} + \epsilon^H_{\mu_j,E}\right).
\end{split}
\end{equation}
In addition, the truncation introduces two additional error probabilities $\epsilon_{M,X}$ and $\epsilon_{M,Z}$. Altogether it follows, from solving the linear programs of Equations~\eqref{eq:LP_Z_finite},~\eqref{eq:LP_E_finite} and~\eqref{eq:LP_X_finite} in Appendix~\ref{ap:finite_vars}, that the expected key rate equals
\begin{equation}
\label{eq:key-rate-LP}
\begin{split}
R = \frac{1-p_{\text{abort}}}{N} \left\lfloor \left( n_{0,1,X}^* - n_X\left(h(e_X)+\delta_{\text{ec}}\left(p_{\text{abort}},n_X\right)\right) \phantom{(\frac{2}{\epsilon_{\text{cor}}(\epsilon_2\epsilon_3(\epsilon_{\text{sec}}-\epsilon))^2}}\right. \right. \\
\qquad \left. \left. -   \log_2\left(\frac{2}{\epsilon_{\text{cor}}(\epsilon_2\epsilon_3(\epsilon_{\text{sec}}-\epsilon))^2}\right) \right) \right\rfloor, 
\end{split}
\end{equation}
where $\epsilon_1,\epsilon_2,\epsilon_3>0$ are arbitrary values such that $\epsilon= 2\epsilon_1 + \epsilon_2+\epsilon_3+\epsilon_e+\epsilon_X+\epsilon_{M,X}+\epsilon_{M,Z}$. 

%% file: Sections/OptimizingKeyRate.tex
\section{Key-Rate Optimization}
\label{sec:model_quantum_channel}
The previous section describes how to compute the key-rate for a given execution of the BB84 protocol. Hence, given the protocol parameters, the number of detection events (Equation~\ref{eq:nr_events}) and the number of errors (Equation~\ref{eq:nr_errors}), the secure key-rate $R$ can be computed by solving three linear programs. 

Next, our objective is to maximize this key rate $R$ by choosing optimal protocol parameters $\mu_j$ and $p_{\mu_j,\basis}$ for $1\leq j\leq m$ and $\basis \in \left\{X,Z\right\}$. To this end, we model the quantum channel such that the expected number of detection events and errors can be computed as function of the protocol parameters.
The physical model is the final step to find the protocol parameter values that optimize secure key-rate $R$. This optimal key-rate may be found by using the following non-linear optimization program: 
\begin{equation}
\label{eq:key_rate_optimization}
\boxed{
	\begin{aligned}
	&&& \max & & R, & \\
	&&& \text{s.t.} & & \forall 0 \leq j \leq m, \basis \in \left\{X,Z\right\} \\
	&&&&& \sum_{j=0}^m p_{\mu_j,X} +p_{\mu_j,Z} =1, \\
	&&&&& \mu_j \geq 0,  \\
	&&&&&  0 \leq p_{\mu_j,\basis} \leq 1.\\
	\end{aligned}
}
\end{equation}
We obtain the solution of this non-linear optimization program by applying the constrained non-linear optimization techniques of~\cite{ConnGT00}.

\subsection{Quantum Channel}
\label{sec:quantum_channel}
We consider a QKD system in which Alice encodes qubits in the polarization of photons and transmits them over a fiber optic cable to Bob. The fiber optic cable is assumed to have an attenuation of $\alpha$~dB/km, i.e., a channel efficiency of $\eta_{ch}= 10^{-\alpha x/10}$ for distance $x$ km~\cite{LCWXZ14}. The channel efficiency equals the fraction of photons that arrive at Bob's detection apparatus, which we assume to be independent of the polarization. In addition to photon losses on the channel, losses can occur in Bob's detection apparatus. These losses are captured by the detector efficiency $\eta_d$. The efficiency of the system with a quantum channel of length $x$ is therefore given by
\begin{equation}
\eta_{sys} = \eta_{ch}\eta_{d} =  10^{-\alpha x/10} \eta_d.
\label{eq:eta_sys}
\end{equation}

Bob's detection apparatus has to be capable of detecting individual photons and is therefore very sensitive. In fact, the photon detectors might click even when they are not illuminated. These events are called dark counts and the probability that a detector clicks without being illuminated is called the dark count probability $p_{dark}$. Recall that Bob's detection apparatus contains two single photon detectors, one for each measurement outcome. Together with the fact that the number of photons in each intensity $\mu_j$ pulse follows a Poisson distribution with mean $\mu_j$, we can derive the following expressing for the gain of  $\mu_j$-pulses in this quantum channel,
\begin{equation}
\label{eq:gain_model}
Q_{\mu_j} = 1- (1-p_{dark})^2 e^{-\mu_j\eta_{sys}}.
\end{equation}
The gain $Q_{\mu_j}$ indicates the fraction of $\mu_j$-pulses that result in a detection event. For a proper derivation of this expression we refer to~\cite{LCWXZ14}. From the gain we easily obtain the expected number of detection events as described in the parameter estimation phase (Equation~\eqref{eq:nr_events}),
\begin{equation}
\label{eq:nr_model}
n_{\mu_j,\basis} = N p_{\mu_j,\basis} Q_{\mu_j}.
\end{equation}
Recall that $N$ is the total number of pulses that is sent and $p_{\mu_j,\basis}$ is the probability that Alice chooses basis $\basis$ and intensity $\mu_j$. 

To model the errors we assume that they are either caused by optical errors in the polarization or by dark counts. Optical errors are modeled by assuming that the polarization of photons is always rotated by an angle $0 \leq \theta \leq \pi/4$. Hence, when Alice sends the qubit $\ket{0}$, Bob receives qubit $\cos(\theta) \ket{0} + \sin(\theta) \ket{0}$. In practice, the polarization error is different per pulse and the angle $\theta$ represents an upper bound on the polarization error. Moreover, $\theta=\pi/4$ results in worst-case behavior, explaining why $\theta$ ranges from $0$ to $\pi/4$. 

Dark counts introduce errors since the associated detection events are independent of the polarization chosen by Alice. Hence, any dark-count event results in an error with probability of $1/2$. Altogether, the expected number of errors for $\mu_j$-pulses in the $\basis$ basis is,
\begin{align}
E_{\mu_j,\basis} = \frac{N p_{\mu_j,\basis}}{2} \left( 1 + (1-p_{dark}) \left( e^{-\mu_j \eta_{sys} \cos^2 \theta} - e^{-\mu_j \eta_{sys} \sin^2 \theta} \right) \right. \nonumber \\
\left.\phantom{e^{-\mu_j \eta_{sys} \cos^2 \theta}} - (1-p_{dark})^2 e^{-\mu_j\eta_{sys}} \right). \label{eq:errors_model}
\end{align}

In this section polarization encoded qubits transmitted over a fiber optic cable were considered. In practice, qubits transmitted over fiber optic cables are often phase encoded~\cite{TRT93}. In contrast, polarization encoding is mostly used to transmit over free space optical communication channels. Considering, these and other quantum channels requires some (minor) modifications to the physical model. 

%% file: Sections/Results.tex
\section{Results}
\label{sec:Results}
In this section we present the results of our key-rate optimization approach. We start by comparing the results from our model to the results presented in~\cite{LCWXZ14}. Afterwards, we consider the effects of increasing the number of pulses sent and increasing the number of intensities used. We denote the number intensity settings by $m$ and number of pulses sent by $N$.

Our experiments comprise three regimes.
As baseline regime we compare our approach using the parameter settings from~\cite{LCWXZ14}. The baseline considers a finite number of intensities $m=3$ and ignores finite key effects $N\rightarrow\infty$ as depicted in Figure~\ref{fig:AsymptoticCompareCurty}. Please note that in contrast to~\cite{LCWXZ14}, we do not fix any of the used intensity settings.
The second regime considers a finite number of intensity settings ($m=3$), and explores the impact of finite key effects by varying the number of pulses $N\in\{10^7,10^8,10^9,10^{10},10^{11}\}$. For comparison, we include our result from the baseline regime (where $N\rightarrow\infty$) as depicted in Figure~\ref{fig:finiteCompare}.
In the third regime we vary the number of intensity settings, while discarding finite key effects, i.e., $N\rightarrow\infty$.
The third regime includes the fully asymptotic case where both $m\rightarrow\infty$ and $N\rightarrow\infty$. This is depicted in Figure~\ref{fig:CompareNumDecoyss}.

\subsection{Baseline}
\label{sec:experiments-baseline}
For the basline regime we adopt the following parameter settings from~\cite{LCWXZ14}:
\begin{itemize}
\item We fix the dark count probability $p_{dark}=6\cdot10^{-7}$ and the detector efficiency $\eta_{d}=0.1$.
\item For the misalignment we take $\theta=0.0707$, with corresponding probability $e_{mis}=5\cdot10^{-3}$.
\end{itemize}
We only compare the results for the asymptotic case, because the results for finite number of pulses are incomparable. In contrast to~\cite{LCWXZ14} we fix the number of pulses sent, while in~\cite{LCWXZ14} the post-processing block-size is fixed. To keep the same post-processing length a higher number of pulses needs to be emitted for larger distances. 
 
\begin{figure}[t]
\centering
	\begin{subfigure}[t]{0.49\textwidth}
	\centering
	\includegraphics[width=\columnwidth]{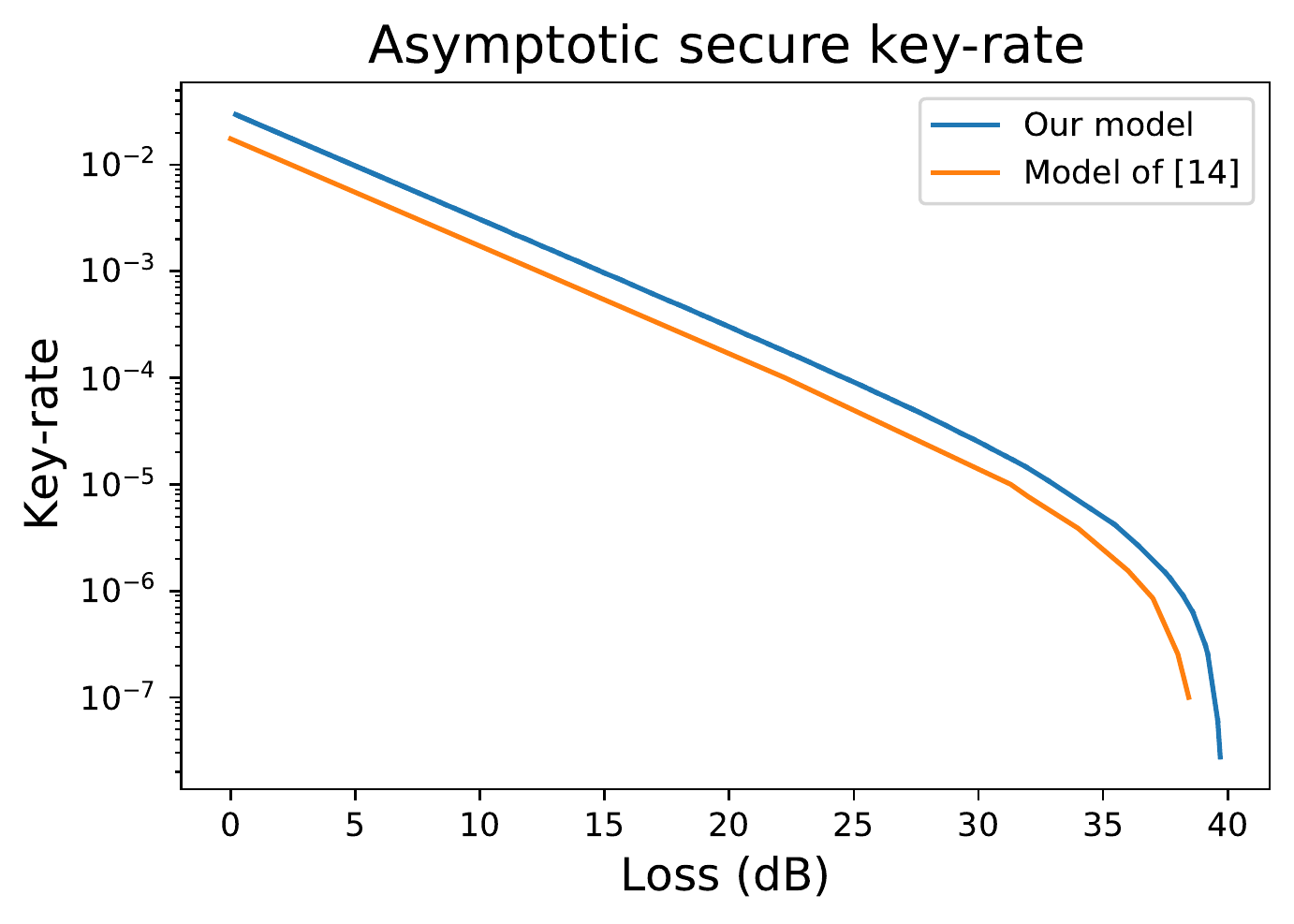}
	\caption{}
	\label{fig:KeyRateAsymptoticCompareCurty}
	\end{subfigure}
	\hfill
	\begin{subfigure}[t]{0.49\textwidth}
	\centering
	\includegraphics[width=\columnwidth]{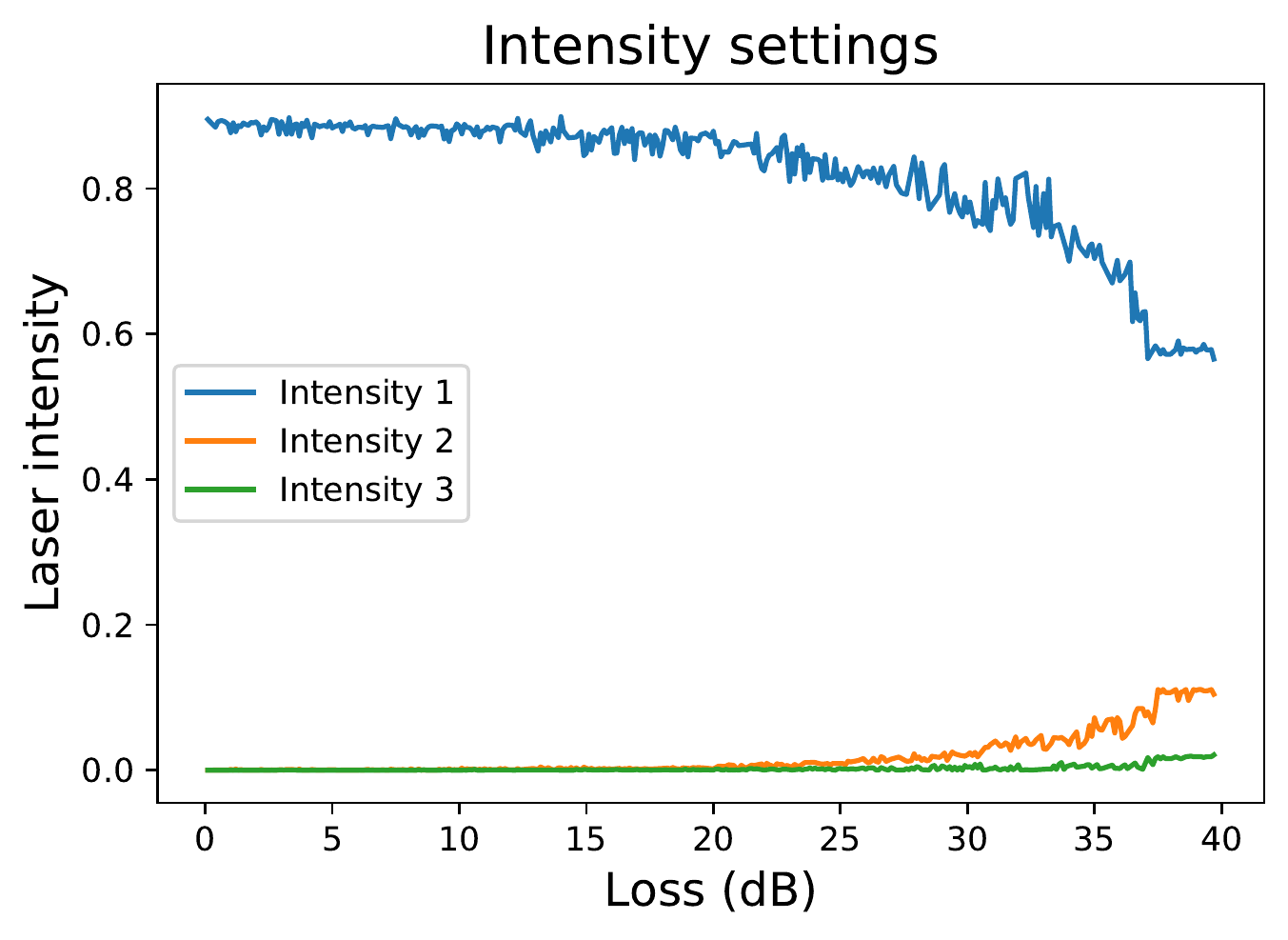}
	\caption{}
	\label{fig:IntensityAsymptoticCompareCurty}
	\end{subfigure}
\caption{The maximum achievable secure key-rate using our model compared to the maximum rate using the model of~\cite{LCWXZ14}. We discard finite key-effects and for our model, we also present the optimal intensity settings per loss.}
\label{fig:AsymptoticCompareCurty}
\end{figure}

Our results for the asymptotic case are shown in Figure~\ref{fig:AsymptoticCompareCurty}. Both the maximum achievable secure key-rate in terms of the loss in decibel (Figure~\ref{fig:KeyRateAsymptoticCompareCurty}) and the optimal intensity settings (Figure~\ref{fig:IntensityAsymptoticCompareCurty}) are presented. The results in~\cite{LCWXZ14} are presented as key-rate per distance in kilometers, however, the two are directly related by an attenuation factor of $0.2$ dB/km. In order to compute a lower bound for the secure key rate we apply the LP's of Section~\ref{sec:unknown}. However these contain infinite sums and assume finite key sizes. We refer to Appendix~\ref{sec:semiAsymptotic} for the LP formulations that discard finite key effects and truncate the infinite sums.
Note that the optimal intensities vary for different channel distances and that, in contrast to~\cite{LCWXZ14}, fixing one of these intensities is sub-optimal. 

With our model the maximum achievable key-rate is higher. However, we did not take into account the after-pulse probability. Depending on the magnitude of this error source, the results of our model may be closer to those of~\cite{LCWXZ14}. It is expected that the used intensities show a rather smooth evolution with increasing distance, however, different behavior is seen. This might result from the optimization routine where a stopping criteria is met too soon, for instance a maximum number of iterations or a local maxima with zero gradient. This can result in a sub-optimal solution and can be overcome by more strict stopping criteria. Despite these artifacts, the found key-rate is higher than obtained in~\cite{LCWXZ14}. Furthermore, secure key material can be extracted for higher losses.

\subsection{Finite-Key Effects}
In this regime we consider the effects of increasing the number of sent pulses $N$ on the key-rate while preserving the baseline settings of Section~\ref{sec:experiments-baseline} including $m=3$. As we include finite key-effects, we have to fix certain cryptographic security parameters. We want to achieve a certain security of our protocol and we want it to be correct with high probability. Therefore, we fix the security and correctness parameters as $\epsilon_{\text{sec}} = \epsilon_{\text{cor}} = 2^{-50}$. Furthermore, we take the abort probability to be $p_{abort}=2^{-50}$. Consequently, we fix $\epsilon_{l,X}^C = \epsilon_{l,Z}^C = \epsilon_{\mu_j,X}^H = \epsilon_{\mu_j,X}^H = \epsilon_{\mu_j,E}^H = 2^{-60}$. This gives upper bounds for $\epsilon_e \le 2^{-54}$ and $\epsilon_X \le 2^{-55}$ and we set $\epsilon_1 =\epsilon_2=\epsilon_3 = 2^{-55}$. Combined this gives $\epsilon \leq 2^{-52}$, which matches with our constraint $\epsilon_{sec} \ge \epsilon$, obtained from Equation~\eqref{eq:key-rate-LP}. 
The linear programs given in Section~\ref{sec:unknown} bound the number of usable pulses and photon errors, but contain an infinite number of variables. We refer to Appendix~\ref{ap:finite_vars} for the truncation of the infinite sums of Section~\ref{sec:unknown}.

For the chosen security parameter $\epsilon_{sec}$ it is sufficient to upper bound multi photon pulses to at most $20$ photons. Indeed, $20$ is a sufficiently large upper bound on the number of photons per pulse: Let $X$ be Poisson distributed with rate $\mu$. According to \cite[Corollary 6]{Short2013}, the Poisson tail probability may be bounded by
\begin{align}
&P(X \geq x \mid \mu=1)\leq \min\left(\frac{e^{-D_{KL}(\mu,x)}}{2},\, \frac{e^{-D_{KL}(\mu,x)}}{\sqrt{2\pi D_{KL}(\mu,x)}}\right),&\text{for }x>\mu,\label{eq:Poisson_Tail_Bound}
\end{align}
where $D_{KL}(\mu,x)$ is the Kullback–Leibler divergence between two Poisson distributed random variables with respective means $\mu$ and $x$:
\begin{equation}
D_{KL}(\mu,x)=\mu-x+x \ln(\frac{x}{\mu}).
\end{equation}
Using the bound from Equation~\eqref{eq:Poisson_Tail_Bound}, we can show that
\begin{equation}
P(X\geq 20\mid \mu=1)<2^{-63.03}<2^{-60}.
\end{equation}

\begin{figure}[!ht]
\centering
\includegraphics[width=0.7\columnwidth]{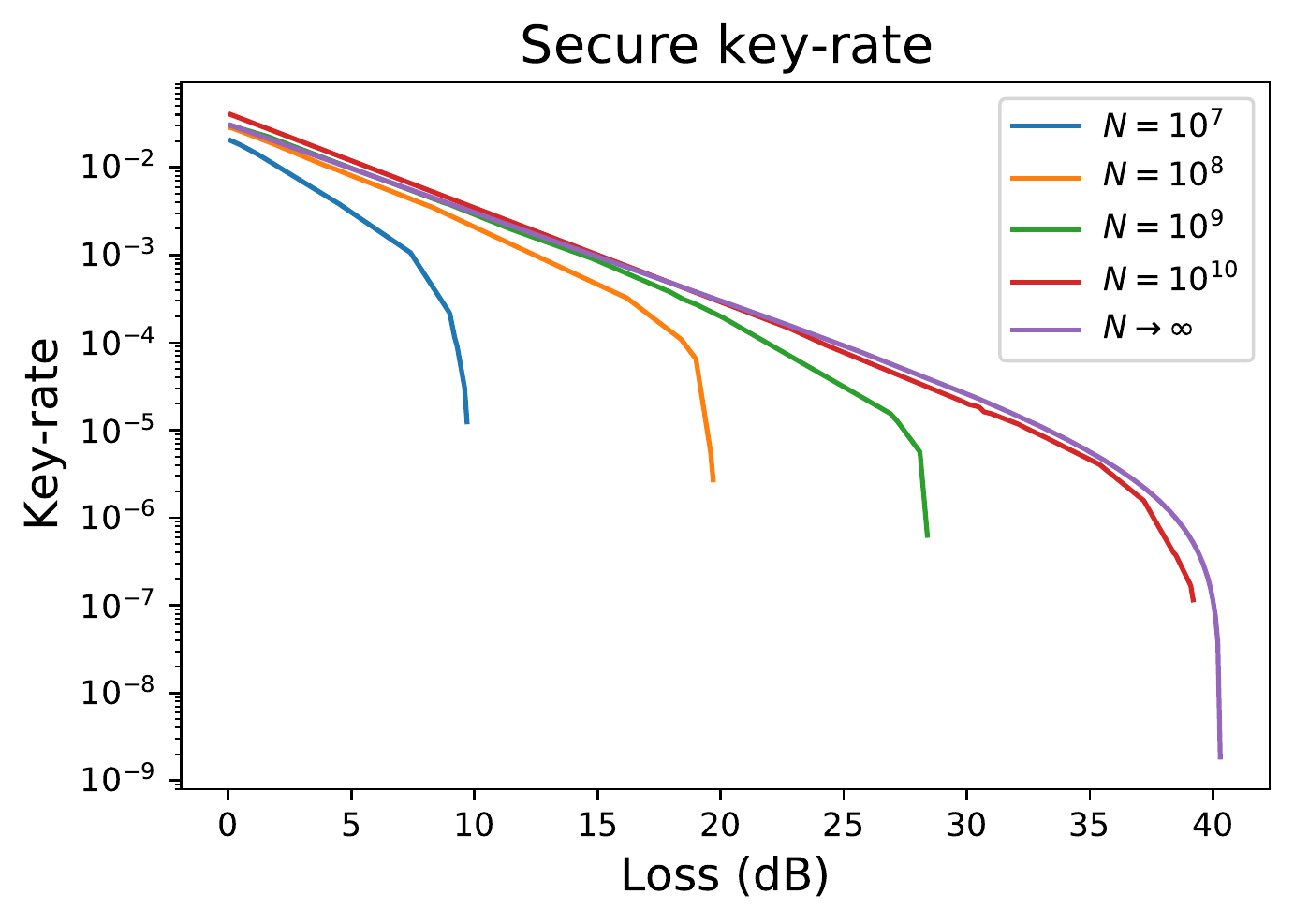}
\caption{The secure key-rate obtained in terms of the losses of the quantum channel (dB). For a varying number of pulses.}
\label{fig:finiteCompare}
\end{figure}

We consider the key-rate for $N = 10^i$ pulses, for $i\in\{7,\hdots,10\}$ and we consider the limit $N\rightarrow\infty$. The results are shown in Figure~\ref{fig:finiteCompare}. The same channel and detector parameters are used as in the baseline. We observe that with increasing number of pulses, the expected secure key-rate indeed increases. Note that already for $10^{10}$ pulses, our key-rate estimation approaches the asymptotic key-rate quite well. The shown figure is the convex hull of the data points. This to account for instabilities in the optimization. We found that for all considered number of pulses sent, the probability that a pulse was sent in the $Z$-basis was less than 6\%, independent of chosen intensity and the loss of the channel. This corresponds with the asymptotic case where the number of pulses sent in the $Z$-basis can be assumed to be an arbitrarily small fraction of the number of pulses. 

\subsection{Asymptotic Key Rates for Different Intensity Settings}
\begin{figure}[!ht]
	\centering
	\begin{subfigure}[t]{0.49\textwidth}
	\centering
	\includegraphics[width=\columnwidth]{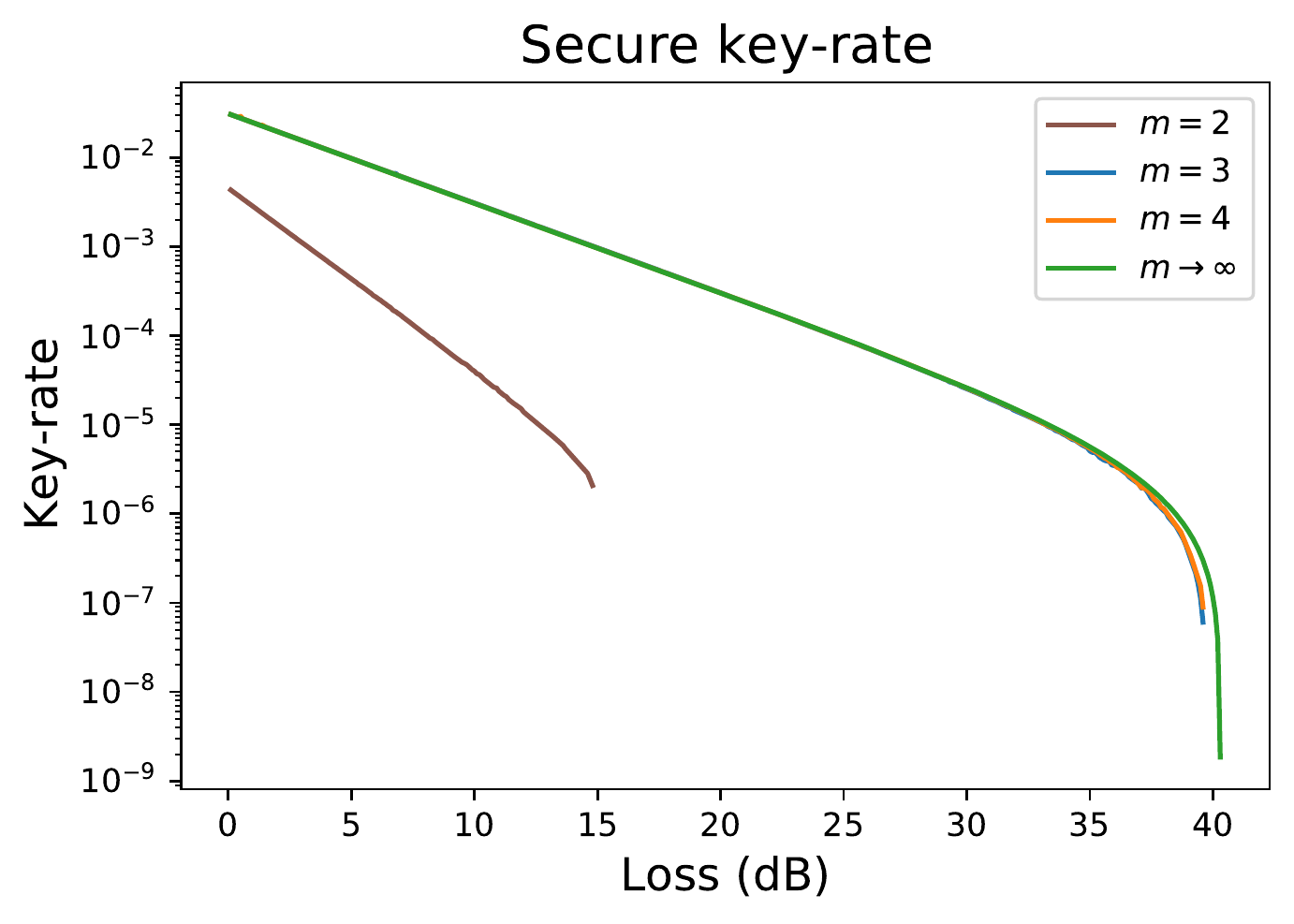}
	\caption{}
	\label{fig:CompareNumDecoys}
	\end{subfigure}
	\hfill
	\begin{subfigure}[t]{0.49\textwidth}
	\centering
	\includegraphics[width=\columnwidth]{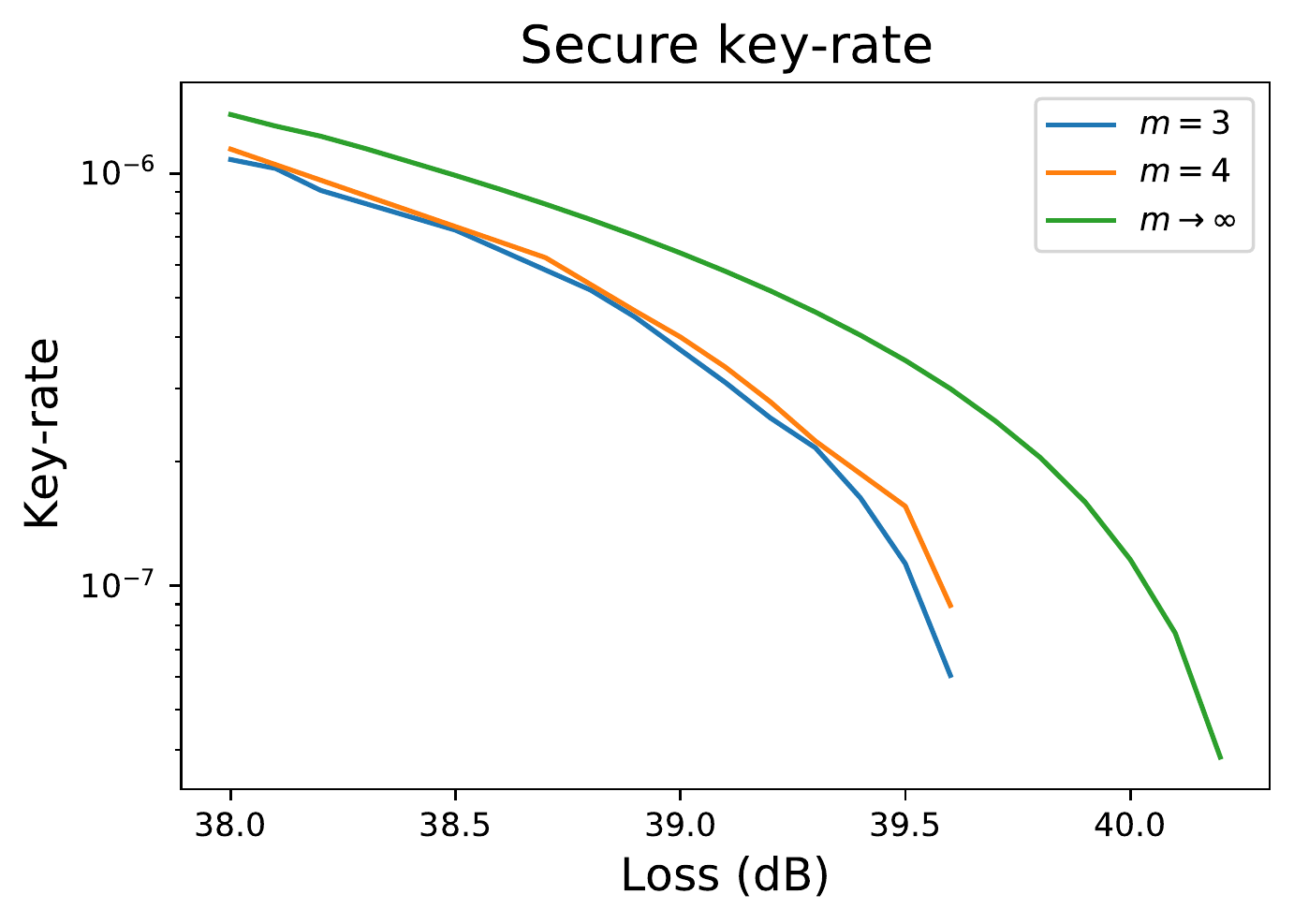}
	\caption{}
	\label{fig:CompareNumDecoysEnd}
	\end{subfigure}
	\caption{The secure key-rate obtained in terms of the loss over the quantum channel (dB) for varying number of intensities $m$.}
	\label{fig:CompareNumDecoyss}
\end{figure}

In this experimental regime we vary the number of intensity settings, for $m\in\{2,3,4\}$ while preserving the baseline settings of Section~\ref{sec:experiments-baseline} including $N\rightarrow\infty$. We also include the fully asymptotic regime, where $m\rightarrow\infty$, and $N\rightarrow\infty$. We refer to Appendix~\ref{sec:fullyAsymptotic} for the unknown parameter estimation in this fully asymptotic case. The key-rate results are presented in Figures~\ref{fig:CompareNumDecoys}~and~\ref{fig:CompareNumDecoysEnd}. We observe that while using only two intensities, the key-rate quickly drops. However, for more than two intensities, the results are very close to each other. Therefore, we focus on regime for $38$ dB up to $40.5$ dB loss in Figure~\ref{fig:CompareNumDecoysEnd}. Here we observe that with each additional intensity, more key-material can be extracted. However, we also see that the gains are marginal. Already for three intensities, losses of up to $39.5$ dB can be tolerated, with key-rate $\sim 10^{-7}$. Using more intensity settings gives only a minor increase in the maximum losses tolerated and a slightly higher key-rate for the same channel lengths. In the limit $m\rightarrow\infty$, the maximum tolerated loss for safely executing the protocol is bounded by $40.1$ dB with a key-rate of about $5\cdot 10^{-8}$. 

%% file: Sections/Conclusion.tex
\section{Conclusion and Discussion}
\label{sec:conclusion}

In this work, we presented a key-rate optimization approach for the decoy-state BB84 QKD protocol. 
Our approach combines several linear and non-linear programs to derive tighter protocol parameters and better key-rates, compared to previous approaches relying on heuristic assumptions. 

Our optimization framework allows the complex optimization problem to be solved, without requiring it to be simplified by means of heuristic assumptions. 
We compared our model to that of~\cite{LCWXZ14} and show that higher key-rates are attained. 
Furthermore, we show the effect of increasing the number of decoy states and we show that using three laser intensities is in general sufficient. 
Thereby validating a heuristic that is commonly used. 

Our work is especially relevant to quantum channels with a significant amount of noise. 
In these cases, the effect of choosing sub-optimal protocol parameters is the largest. 
Some settings do not even allow any key material to extracted when sub-optimal protocol parameters are used. 
In particular, our parameter settings allow for higher losses to be tolerated.

The analysis of this work focused on the BB84 QKD protocol. 
However, similar analyses can also be applied to other QKD protocols, such as for instance BBM92 or measurement-device independent protocols. 
The model can also be extended to incorporate more practical disturbances and noise. Furthermore, the model can be used in practical settings to optimize QKD protocol parameters to obtain higher key-rates.

%% file: Sections/Appendix.tex
\section{Notation}

\begin{table}[H]
	\centering
	\begin{tabular}{|c|c|c|}
		\hline
		Variable & Domain & Description  \\ \hline
		$N$ & $\mathbb{Z}_{\geq 0}$ & Number of pulses sent by Alice \\ \hline
		$\basis$ & $\{X,Z\}$ & Basis \\ \hline
		$U$ & $\{\mu_0,\dots,\mu_m\}$ & Intensities \\ \hline
		$n$ & $\mathbb{Z}_{\geq 0}$ & Number of pulses in which Alice and Bob chose \\ &&  the same basis and Bob  registered an detection \\ \hline
		$E$ & $\mathbb{Z}_{\geq 0}$ & Number of errors \\ \hline
		$e$ & $\left[0,1/2\right]$ & Error rate \\ \hline
		$\mu$ & $\mathbb{R}_{\geq 0}$ & Pulse intensity (mean photon number)  \\ \hline
		$p$ & $\left[0,1\right] $ & Probability \\ \hline
		$\ell$ & $\mathbb{Z}_{\geq 0}$ & Length of the secret key (in bits) \\ \hline
		$R$ & $\left[0,1\right] $ & Secret key rate \\ \hline
		$K^{r}$ & $\{0,1\}^N$ & Raw key \\ \hline
		$K^{s}$ & $\{0,1\}^{n_s}$ & Sifted key \\ \hline
		$K^{e}$ & $\{0,1\}^{n_X}$ & Error corrected key \\ \hline
		$K$ & $\{0,1\}^{\ell}$ & Secure key \\ \hline
	\end{tabular}
	\caption{Overview of the most important variables. Subscripts are used to condition on events such as basis, intensity or photon number. Superscripts indicate whether the key is Alice's ($a$) or Bob's ($b$)}
\end{table}

\section{Linear programs with finite number of variables}
\label{ap:finite_vars}

The linear programs given in Section~\ref{sec:unknown} contain an infinite number of variables. To reduce the number of variables we truncate the infinite sums at $M$ and upper-bound the number of pulses containing more than $M$ photons, 
\begin{equation}
m_{M,\basis} = \sum_{l=M+1}^{\infty} n_{l,\basis}. 
\end{equation}
Namely note that, 
\begin{equation}
e^{-\mu_j}p_{\mu_j|\basis} \sum_{l=0}^{M} \frac{\mu_j^{l}}{l!}\frac{n_{l,\basis}}{p_{l|\basis}} \leq e^{-\mu_j}p_{\mu_j|\basis} \sum_{l=0}^{\infty} \frac{\mu_j^{l}}{l!}\frac{n_{l,\basis}}{p_{l|\basis}} \leq m_{M,\basis} + e^{-\mu_j}p_{\mu_j|\basis} \sum_{l=0}^{M} \frac{\mu_j^{l}}{l!}\frac{n_{l,\basis}}{p_{l|\basis}}.
\end{equation}
Hence, these inequalities allow us to bound the infinite sum by two finite sums. 

Let $q_{M,\basis}$ be the probability that a $\basis$-pulse contains more than $M$ photons, i.e. 
\begin{equation}
q_{M,\basis} = \sum_{j=0}^m p_{\mu_j|\basis}  e^{-\mu_j} \sum_{l=M+1}^{\infty} \frac{\mu_j^{l}}{l!}.
\end{equation}  
For
\begin{equation}
\Lambda_{M,\basis} : = q_{M,\basis}N_\basis+f\left(N_\basis;q_{M,\basis};\epsilon_{M,\basis}\right),
\end{equation}
it follows, by Chernoff's bound, that 
\begin{equation}
\Prob\left( m_{M,\basis} \geq \Lambda_{M,\basis} \right) \leq \epsilon_{M,\basis}, 
\end{equation}
for all $\epsilon_{M,\basis}>0$. Hence solving the linear programs of Equations~\eqref{eq:LP_Z_finite},~\eqref{eq:LP_E_finite} and~\eqref{eq:LP_X_finite}, that contain a finite number of variables, will allow us to compute the length $\ell^*$ of the secret key, 
\begin{equation}
\ell^* := n_{0,1,X}^* - n_X\left(h(e_X)+\delta_{\text{ec}}\left(p_{\text{abort}},n_X\right)\right)  -   \log_2\left(\frac{2}{\epsilon_{\text{cor}}(\epsilon_2\epsilon_3(\epsilon_{\text{sec}}-\epsilon))^2}\right),
\label{eq:keyRateFunctionFinite}
\end{equation}
for some $\epsilon_1,\epsilon_2,\epsilon_3>0$ and $\epsilon= 2\epsilon_1 + \epsilon_2+\epsilon_3+\epsilon_e+\epsilon_X + \epsilon_{M,X} +\epsilon_{M,Z} $.

\begin{equation}
\label{eq:LP_Z_finite}
\boxed{
\begin{aligned}
n_{1,Z}^*  = &&& \min & & n_{1,Z}, & \\
&&& \text{s.t.} & &  \forall \, \leq j \leq m, \forall\, 0 \leq l \leq M \\
&&&&& n_{l,Z}, \delta_{\mu_j} \in \mathbb{R}, \\
&&&&& n_{\mu_j,Z} + \delta_{\mu_j} \geq e^{-\mu_j}p_{\mu_j|Z} \sum_{l=0}^{M} \frac{\mu_j^{l}}{l!}\frac{n_{l,Z}}{p_{l|Z}}, \\
&&&&& n_{\mu_j,Z} + \delta_{\mu_j} \leq e^{-\mu_j}p_{\mu_j|Z} \sum_{l=0}^{M} \frac{\mu_j^{l}}{l!}\frac{n_{l,Z}}{p_{l|Z}} + \Lambda_{M,Z}, \\
&&&&&  \sum_{j=0}^m \delta_{\mu_j} =0, \\
&&&&& 0 \leq n_{l,Z} \leq \min\left(p_{l|Z}N_Z+  f(N_Z;p_{l|Z};\epsilon^C_{l,Z}), n_Z \right), \\ 
&&&&&  \left\lvert \delta_{\mu_j} \right\rvert \leq \sqrt{-\ln(\epsilon^H_{\mu_j,Z}/2)n_Z/2}.
\end{aligned}
}
\end{equation}

\begin{equation}
\label{eq:LP_E_finite}
\boxed{
\begin{aligned}
E_{1,Z}^* = &&& \max & & E_{1,Z} & \\
&&& \text{s.t.} & & \forall \,0 \leq j \leq m, \forall \,0 \leq l \leq M \\
&&&&& E_{l,Z}, \delta_{\mu_j} \in \mathbb{R}, \\
&&&&& E_{\mu_j,Z} + \delta_{\mu_j} \geq e^{-\mu_j}p_{\mu_j|Z} \sum_{l=0}^{M} \frac{\mu_j^{l}}{l!} \frac{E_{l,Z}}{p_{l|Z}}, \\
&&&&& E_{\mu_j,Z} + \delta_{\mu_j} \leq e^{-\mu_j}p_{\mu_j|Z} \sum_{l=0}^{M} \frac{\mu_j^{l}}{l!} \frac{E_{l,Z}}{p_{l|Z}} + \Lambda_{M,Z}, \\
&&&&&  \sum_{j=0}^m \delta_{\mu_j} =0, \\
&&&&& 0 \leq E_{l,Z} \leq \min\left(p_{l|Z}N_{Z} +  f(N_Z;p_{l|Z};\epsilon^C_{l,Z}) , E_Z \right), \\
&&&&&  \left\lvert \delta_{\mu_j} \right\rvert \leq \sqrt{-\ln(\epsilon^H_{\mu_j,e}/2)E_Z/2},  \\
\end{aligned}
}
\end{equation}

\begin{equation}
\label{eq:LP_X_finite}
\boxed{
\begin{aligned}
n_{0,1,X}^* = &&& \min & & n_{0,X}+n_{1,X}-n_{1,X}h(e_{1,Z}^*+\delta(n_X,n_Z,\epsilon_1)), & \\
&&& \text{s.t.} & & \forall \,0 \leq j \leq m, \forall \,0 \leq l \leq M \\
&&&&& n_{l,X}, \delta_{\mu_j} \in \mathbb{R}, \\
&&&&& n_{\mu_j,X} + \delta_{\mu_j} \geq e^{-\mu_j}p_{\mu_j|X} \sum_{l=0}^{M} \frac{\mu_j^{l}}{l!} \frac{n_{l,X}}{p_{l|X}},  \\
&&&&& n_{\mu_j,X} + \delta_{\mu_j} \leq e^{-\mu_j}p_{\mu_j|X} \sum_{l=0}^{M} \frac{\mu_j^{l}}{l!} \frac{n_{l,X}}{p_{l|X}} + \Lambda_{M,X},  \\ 
&&&&&  \sum_{j=0}^m \delta_{\mu_j} =0, \\
&&&&& 0 \leq n_{l,X} \leq  \min\left(p_{l|Z}N_X+\ f(N_X;p_{l|X};\epsilon^C_{l,X}), n_X \right), \\
&&&&&  \left\lvert \delta_{\mu_j} \right\rvert \leq \sqrt{-\ln(\epsilon^H_{\mu_j,X}/2)n_X/2}.
\end{aligned}
}
\end{equation}

\section{Asymptotic case}

In the asymptotic limit, i.e. when $N\to \infty$, the Serfling, Hoeffding and Chernoff terms vanish and the linear programs simplify significantly. In this section the asymptotic linear programs are presented. First, the case with a finite amount of decoy states is presented. Subsequently, we consider the case where the number of decoy intensities $m$ goes to infinity as well. In this case the linear programs can be omitted entirely. 

\subsection{Finite number of decoy intensities}
\label{sec:semiAsymptotic}
Let us first define the yields $Y_{l,\basis}$ and the gains $Q_{\mu_j,\basis}$, 

\begin{equation}
\begin{split}
Y_{l,\basis} & : = \frac{n_{l,\basis}}{p_{l|\basis}N_{\basis}}, \quad \forall l \geq 0,\basis \in\{X,Z\}, \\
Q_{\mu_j,\basis} & : = \frac{n_{\mu_j,\basis}}{p_{\mu_j|\basis}N_{\basis}}, \quad  \forall 0\leq j \leq m,\basis \in\{X,Z\}. 
\end{split}
\end{equation}

In addition, we define the following variables,
\begin{equation}
\begin{split}
\gamma_{l,Z} & := e_{l,Z}Y_{l,Z} = \frac{E_{l,Z}Y_{l,Z}}{n_{l,Z}} = \frac{E_{l,Z}}{p_{l|Z}N_{Z}}, \quad \forall l \geq 0, \\
\gamma_{\mu_j,Z} & := e_{\mu_j,Z}Q_{\mu_j,Z} = 
\frac{E_{\mu_j,Z}Q_{\mu_j,Z}}{n_{\mu_j,Z}} = 
\frac{E_{\mu_j,Z}}{p_{\mu_j|Z}N_{Z}}, \quad  \forall 0\leq j \leq m. 
\end{split}
\end{equation}

Substituting the above variables in Equations~\eqref{eq:LP_Z},~\eqref{eq:LP_E} and~\eqref{eq:LP_X} results in the following linear programs, where $e_{1,Z}^* := \frac{\gamma_{1,Z}^*}{Y_{1,Z}^*}$.

\begin{equation}
\boxed{
\begin{aligned}
Y_{1,Z}^* = &&& \min & & Y_{1,Z}, & \\
&&& \text{s.t.} & & \forall \,0 \leq j \leq m, \forall \, l\geq 0 \\
&&&&& Q_{\mu_j,Z} = e^{-\mu_j}\sum_{l=0}^{\infty} \frac{\mu_j^{l}}{l!} Y_{l,Z}, \\
&&&&& 0 \leq Y_{l,Z} \leq 1.
\end{aligned}
}
\end{equation}

\begin{equation}
\boxed{
\begin{aligned}
\gamma_{1,Z}^* = &&& \max & & \gamma_{1,Z} & \\
&&& \text{s.t.} & & \forall \,0 \leq j \leq m, \forall \, l\geq 0 \\
&&&&& \gamma_{\mu_j,Z} = e^{-\mu_j} \sum_{l=0}^{\infty} \frac{\mu_j^{l}}{l!} \gamma_{l,Z}, \\
&&&&& 0 \leq \gamma_{l,Z} \leq 1.
\end{aligned}
}
\end{equation}

\begin{equation}
\label{LP_01:asymp}
\boxed{
\begin{aligned}
Y_{0,1,X}^* = &&& \min & & p_{0|X} Y_{0,X}+p_{1|X} Y_{1,X}\left(1-h(e_{1,Z}^*)\right), & \\
&&& \text{s.t.} & & \forall \,0 \leq j \leq m, \forall \, l\geq 0 \\
&&&&&  Q_{\mu_j,X} = e^{-\mu_j} \sum_{l=0}^{\infty} \frac{\mu_j^{l}}{l!} Y_{l,X}, \\
&&&&& 0 \leq Y_{l,X} \leq 1.
\end{aligned}
}
\end{equation}

Solving these linear programs results in a secure key-rate 
\begin{equation}
R^* = Y_{0,1,X}^*  - Q_X h(e_X). 
\label{eq:finiteDecoyFunction}
\end{equation}

In linear program~\ref{LP_01:asymp}, we have used the fact that the sifting probability $p_X$ can be taken arbitrarily close to $1$. 

\subsection{Infinite number of decoy intensities}
\label{sec:fullyAsymptotic}

If, in addition, we assume an infinite amount of decoy intensities (i.e. $m \to \infty$) then, for properly chosen intensities, the linear programs can be shown to posses a single feasible solution. Hence, Alice and Bob can in this case compute the exact yields and error rates. The resulting key rate can therefore be computed as follows,
\begin{equation}
R^* = p_{0|X} Y_{0,X}+p_{1|X}Y_{1,X}\left(1-h(e_{1,Z})\right)  - Q_X h(e_X). 
\label{eq:bothInfiniteFunction}
\end{equation}

The sifting probabilities $p_{0|X}$ and $p_{1|X}$ depend on the intensities, which are chosen to maximize the key rate.

%% file: main.bbl
\begin{thebibliography}{10}

\bibitem{BB84}
C.~H. Bennett and G.~Brassard, ``An update on quantum cryptography,'' in {\em
  Advances in Cryptology, Proceedings of {CRYPTO} '84, Santa Barbara,
  California, USA, August 19-22, 1984, Proceedings} (G.~R. Blakley and
  D.~Chaum, eds.), vol.~196 of {\em Lecture Notes in Computer Science},
  pp.~475--480, Springer, 1984.

\bibitem{GRTZ02}
N.~{Gisin}, G.~{Ribordy}, W.~{Tittel}, and H.~{Zbinden}, ``{Quantum
  cryptography},'' {\em Reviews of Modern Physics}, vol.~74, pp.~145--195, Jan
  2002.

\bibitem{May96}
D.~Mayers, ``Quantum key distribution and string oblivious transfer in noisy
  channels,'' in {\em Advances in Cryptology - {CRYPTO} '96, 16th Annual
  International Cryptology Conference, Santa Barbara, California, USA, August
  18-22, 1996, Proceedings} (N.~Koblitz, ed.), vol.~1109 of {\em Lecture Notes
  in Computer Science}, pp.~343--357, Springer, 1996.

\bibitem{BHLMO05}
M.~Ben{-}Or, M.~Horodecki, D.~W. Leung, D.~Mayers, and J.~Oppenheim, ``The
  universal composable security of quantum key distribution,'' in {\em Theory
  of Cryptography, Second Theory of Cryptography Conference, {TCC} 2005,
  Cambridge, MA, USA, February 10-12, 2005, Proceedings} (J.~Kilian, ed.),
  vol.~3378 of {\em Lecture Notes in Computer Science}, pp.~386--406, Springer,
  2005.

\bibitem{Can01}
R.~Canetti, ``Universally composable security: {A} new paradigm for
  cryptographic protocols,'' in {\em 42nd Annual Symposium on Foundations of
  Computer Science, {FOCS} 2001, 14-17 October 2001, Las Vegas, Nevada, {USA}},
  pp.~136--145, {IEEE} Computer Society, 2001.

\bibitem{XMZLP20}
F.~Xu, X.~Ma, Q.~Zhang, H.-K. Lo, and J.-W. Pan, ``Secure quantum key
  distribution with realistic devices,'' {\em Rev. Mod. Phys.}, vol.~92,
  p.~025002, May 2020.

\bibitem{BLMS00b}
G.~Brassard, N.~L{\"{u}}tkenhaus, T.~Mor, and B.~C. Sanders, ``Security aspects
  of practical quantum cryptography,'' in {\em Advances in Cryptology -
  {EUROCRYPT} 2000, International Conference on the Theory and Application of
  Cryptographic Techniques, Bruges, Belgium, May 14-18, 2000, Proceeding}
  (B.~Preneel, ed.), vol.~1807 of {\em Lecture Notes in Computer Science},
  pp.~289--299, Springer, 2000.

\bibitem{Lut00}
N.~L\"utkenhaus, ``Security against individual attacks for realistic quantum
  key distribution,'' {\em Phys. Rev. A}, vol.~61, p.~052304, Apr 2000.

\bibitem{Hwa03}
W.-Y. Hwang, ``Quantum key distribution with high loss: Toward global secure
  communication,'' {\em Phys. Rev. Lett.}, vol.~91, p.~057901, Aug 2003.

\bibitem{LMC05}
H.-K. Lo, X.~Ma, and K.~Chen, ``Decoy state quantum key distribution,'' {\em
  Phys. Rev. Lett.}, vol.~94, p.~230504, Jun 2005.

\bibitem{Wan05}
X.-B. Wang, ``Beating the photon-number-splitting attack in practical quantum
  cryptography,'' {\em Phys. Rev. Lett.}, vol.~94, p.~230503, Jun 2005.

\bibitem{TR11}
M.~Tomamichel and R.~Renner, ``Uncertainty relation for smooth entropies,''
  {\em Phys. Rev. Lett.}, vol.~106, p.~110506, Mar 2011.

\bibitem{TLGR11}
M.~Tomamichel, C.~C.~W. Lim, N.~Gisin, and R.~Renner, ``Tight finite-key
  analysis for quantum cryptography,'' {\em Nature Communications}, vol.~3, Jan
  2012.

\bibitem{LCWXZ14}
C.~C.~W. Lim, M.~Curty, N.~Walenta, F.~Xu, and H.~Zbinden, ``Concise security
  bounds for practical decoy-state quantum key distribution,'' {\em Phys. Rev.
  A}, vol.~89, Feb 2014.

\bibitem{MQZL05}
X.~Ma, B.~Qi, Y.~Zhao, and H.-K. Lo, ``Practical decoy state for quantum key
  distribution,'' {\em Phys. Rev. A}, vol.~72, Jul 2005.

\bibitem{ConnGT00}
A.~R. Conn, N.~I.~M. Gould, and P.~L. Toint, {\em Trust Region Methods}.
\newblock {MOS-SIAM} Series on Optimization, {SIAM}, 2000.

\bibitem{LPD13}
M.~Lucamarini, K.~A. Patel, J.~F. Dynes, B.~Fr\"{o}hlich, A.~W. Sharpe, A.~R.
  Dixon, Z.~L. Yuan, R.~V. Penty, and A.~J. Shields, ``Efficient decoy-state
  quantum key distribution with quantified security,'' {\em Opt. Express},
  vol.~21, pp.~24550--24565, Oct 2013.

\bibitem{CXC14}
M.~Curty, F.~Xu, W.~Cui, C.~C.~W. Lim, K.~Tamaki, and H.-K. Lo, ``Finite-key
  analysis for measurement-device-independent quantum key distribution,'' {\em
  Nature Communications}, vol.~5, Apr. 2014.

\bibitem{CW79}
L.~Carter and M.~N. Wegman, ``Universal classes of hash functions,'' {\em J.
  Comput. Syst. Sci.}, vol.~18, no.~2, pp.~143--154, 1979.

\bibitem{Renner05}
R.~Renner, ``Security of quantum key distribution,'' in {\em Ausgezeichnete
  Informatikdissertationen 2005} (D.~Wagner, ed.), vol.~{D-6} of {\em {LNI}},
  pp.~125--134, {GI}, 2005.

\bibitem{TSSR11}
M.~Tomamichel, C.~Schaffner, A.~D. Smith, and R.~Renner, ``Leftover hashing
  against quantum side information,'' {\em {IEEE} Trans. Information Theory},
  vol.~57, no.~8, pp.~5524--5535, 2011.

\bibitem{VDTR13}
A.~Vitanov, F.~Dupuis, M.~Tomamichel, and R.~Renner, ``Chain rules for smooth
  min- and max-entropies,'' {\em {IEEE} Trans. Information Theory}, vol.~59,
  no.~5, pp.~2603--2612, 2013.

\bibitem{Serfling1974}
R.~J. Serfling, ``Probability inequalities for the sum in sampling without
  replacement,'' {\em Ann. Statist.}, vol.~2, pp.~39--48, 01 1974.

\bibitem{Hoeffding63}
W.~Hoeffding, ``Probability inequalities for sums of bounded random
  variables,'' {\em Journal of the American Statistical Association}, vol.~58,
  no.~301, pp.~13--30, 1963.

\bibitem{Chernoff1952}
H.~Chernoff, ``A measure of asymptotic efficiency for tests of a hypothesis
  based on the sum of observations,'' {\em Ann. Math. Statist.}, vol.~23,
  pp.~493--507, 12 1952.

\bibitem{TRT93}
P.~D. {Townsend}, J.~G. {Rarity}, and P.~R. {Tapster}, ``Single photon
  interference in 10 km long optical fibre interferometer,'' {\em Electronics
  Letters}, vol.~29, pp.~634--635, April 1993.

\bibitem{Short2013}
M.~Short, ``Improved inequalities for the poisson and binomial distribution and
  upper tail quantile functions,'' {\em ISRN Probability and Statistics},
  vol.~2013, 2013.

\end{thebibliography}
